\documentclass[12pt]{article}

\usepackage{amsmath,amsthm,amssymb}
\usepackage{graphics,epsfig}
\begin{document}
	
	\begin{center}
		\large{\bf{Motion of charged particle in  Reissner - Nordstr\"{o}m  spacetime:\\
	A Jacobi metric  approach}}	
	\end{center}
	\vspace{0.0cm}
	\begin{center}
		Praloy Das \footnote{E-mail: praloydasdurgapur@gmail.com}, Ripon Sk \footnote{E-mail: riponphysics@gmail.com} and
		Subir Ghosh \footnote{E-mail: subirghosh20@gmail.com}\\
		\vspace{2.0 mm}
		\small{\emph{Physics and Applied Mathematics Unit, Indian Statistical
				Institute\\
				203 B. T. Road, Kolkata 700108, India}} \\
	\end{center}
	\vspace{0.5cm}
	\vskip 1cm

\begin{abstract}
The present work discusses motion of neutral and charged particles in Reissner - Nordstr\"{o}m  spacetime. The constant energy paths are derived in a variational principle framework using the Jacobi metric which is parameterized by conserved particle energy. Of particular interest is the case of particle charge and Reissner-Nordstr\"{o}m  black hole charge being of same sign since this leads to a clash of opposing forces - gravitational (attractive) and Coulomb (repulsive). Our paper aims to compliment the recent works of Pugliese,  Quevedo and Ruffini [1,2]. The energy dependent Gaussian curvature (induced by Jacobi metric), plays an important role in classifying the trajectories.	
\end{abstract}
\section{Introduction}

Point particle dynamics serves the important purpose of initiating the study more complex motion of extended bodies. A major area of interest, especially in recent times, is the motion of particles in the presence of black holes and naked singularities. The exhaustive topical works by Pugliese,  Quevedo and Ruffini \cite{p1,p2} (as well as their series of earlier papers \cite{p3}) have revealed clearly the possibility of distinguishing between black holes and naked singularities in the case of charged particle dynamics in Reissner - Nordstr\"{o}m background. The systematic analysis is based on the nature of particle motion for different charge to mass ratio in the Reissner - Nordstr\"{o}m metric, (which is expected), but the non-trivial effect of the charge to mass ratio of the probe particle is indeed a surprising element. In this perspective, the aim of the present paper, dealing once again on the motion of point charge in the presence of a charged black hole, is not to add any more detail on the results of \cite{p1,p2, gru}, but to compliment their observations and bring in new insights from an entirely different, (and hitherto less explored), point of view, that of Jacobi metric approach. This framework has generated a lot of interest after the series of works by Gibbons \cite{gib} and by Chanda,  Gibbons and Guha \cite{gibb}.

The primary major difference is the following. On the one hand,  Pugliese et.al. \cite{p1,p2,p3} relies on the covariant framework where geodesic motion takes place in Reissner-Nordstr\"{o}m spacetime. The Lagrangian equation of motion is derived from the unrestricted variational principle with the charge and mass parameters of the source and probe dictating the particle motion for arbitrary particle energy, which is a derived quantity. On the other hand, in Jacobi metric framework, {\it{the equation of motion is obtained from a restricted variational principle with fixed energy}} where the Lagrangian depends on the Jacobi metric that involves the particle energy explicitly. Schematically for a system with $E=K-V$ where $E,K,V$ are total energy, kinetic energy and potential energy of a probe particle respectively, the Jacobi metric is given by 
\begin{equation}
	\label{001}
	j=2(E-V)g
\end{equation}
where $j$ is the Jacobi metric and  $g$ the Riemannian metric on the manifold where the motion is taking place.  In a sense each particle is performing geodesic motion in its own space(time). The Jacobi metric is manifestly non-covariant (on a spatial slice) \cite{gib,gibb} and the equation of motion is derived from a restricted (Maupertuis) variational principle with fixed particle energy. Hence energy acts as an additional parameter along with the charge and mass parameters. The particle motion is still geodesic in the Jacobi metric space. (The fact that Jacobi metric is degenerate at the boundary where $E=V$ can create complications. For a detailed analysis on this issue and  on ways of overcoming it, see for example \cite{szy}.)

A  novelty in our analysis  is the associated (Gaussian) curvature corresponding to the Jacobi metric, in line with \cite{gib}.   In the next section we will see that even dynamics in flat space can give rise to an effective non-zero curvature from Jacobi metric space. In fact trajectories can be classified according to the sign of the curvature: positive, negative and zero curvature (for Jacobi metric) corresponds to elliptic (with negative energy), hyperbolic (positive energy) and parabolic (zero energy) orbits respectively.

Apart from the importance of curvature in the Jacobi metric, one way in which the present work can compliment the exhaustive analysis of \cite{p1,p2} is the following: whereas \cite{p1,p2} relies on extremely detailed graphical analysis of the orbit structure keeping {\it{exact}} expressions we have provided {\it{analytic expressions}} of the orbits, albeit in a perturbative framework (of small charge of both the black hole and the probe). Also, compared to \cite{p1,p2} we have not restricted the analysis to circular orbits only.

We conclude this brief introduction by mentioning structure of the present work. The paper is organized as follows: In Section 2 we discuss generalities related to Maupertuis principle and Jacobi metric as well as its application in problems of particle dynamics in black hole spacetimes. Section 3 deals with the neutral massive particle motion in  Reissner - Nordstr\"{o}m  background. Section 4 constitutes the study of motion of charged massive particle in Reissner - Nordstr\"{o}m background. In Section 5 we provide a detailed analysis of Gaussian curvature for Jacobi metrics in the cases under study. The paper ends with our conclusions in Section 6. Detailed results are given Appendices 7 and 8.
\section{Maupertuis principle and Jacobi metric}

The Maupertuis transformation and the associated Jacobi metric approach has generated a lot of recent interest \cite{gib,gibb}. To understand the context let us start by considering an  obvious fact  that a curve can be parameterized in infinite number of ways (see for example \cite{sig}  for a detailed discussion). For example, an ellipse

 \begin{equation}
 	\label{01}
 	\frac{x^2}{a^2}+\frac{y^2}{b^2}=1
 	\end{equation}
 can be parameterized by 	
 	\begin{equation}
 x=a~sin(t),~~y=b~cos(t),	
\label{02}
\end{equation}
and also by a different parameter $\tilde t$ related to $t$ by 
\begin{equation}
\label{03}
\tilde t=(a^2+b^2)t+(a^2-b^2)sin(t).
\end{equation}
Clearly the former parameterization (\ref{02}) is connected to a two dimensional oscillator whereas the latter (\ref{03}) is related to the Kepler problem. Generically the same trajectory can be attributed to distinct integrable systems where the parameter $t$ plays the role of time variable,  conjugate to the respective Hamiltonian.  Hence the question is how to derive the particular parameterization that matches with a known or interesting problem. The Maupertuis principle recasts this problem in a dynamical setup. The Maupertuis variational principle states that, given an $n$-dimensional configuration space Lagrangian $L(q_i,\dot q_i)$, extremals of the action
\begin{equation}
\label{04}
S=\int dt~L(q_i,\dot q_i)
\end{equation}
coincide with extremals of the reduced action in a $2n-1$-dimensional phase space that is a level set of the Hamiltonian function $H(q_i,p_i)=E$. Here $p_i$ constitute the conjugate momenta to $q_i$ and the variation takes place on a fixed energy ($E$) hypersurface. In a more explicit way let us suppose a natural Hamiltonian in a Riemanniam manifold with metric $g_{ij}$,
\begin{equation}
\label{05}
H(q,p)=T(q,p)+V(q)=g_{ij}(q)p_ip_j+V(q).
\end{equation}
The trajectories for $H$ on a  fixed energy smooth submanifold will coincide with trajectories of a new Hamiltonian $\tilde H$ given by
\begin{equation}
\tilde H(q,p)=\tilde T(q,p)=\frac{g_{ij}(q)}{E-V(q)}p_ip_j,
\label{06}
\end{equation}
along with a transformation of the parameters,
\begin{equation}
\label{07}
d\tilde t=(E-V(q))dt.
\end{equation}
Together (\ref{06},\ref{07}) constitute the Jacobi transformation. A scaling of the effective metric provides the Jacobi metric. The two major novelties and advantages of the Maupertuis-Jacobi framework are:\\
(i) The Hamiltonian $\tilde H$ yields geodesic motion. This allows one to treat a dynamical problem as geodesic motion and well known machinery of geodesic motion can be directly exploited. \\ (ii)   The Maupertuis principle and subsequent Jacobi transformation preserve integrability that is under this map an integrable system with a natural Hamiltonian goes over to another integrable system in the same phase space. Thus this scheme provides a method to search for new integrable systems.\\
(iii) The Gaussian curvature induced by Jacobi metric allows one to classify the trajectories based on their energy since it appears explicitly in Jacobi metric.

For a generic Lagrangian of the form, (following the notation of \cite{gib})
\begin{eqnarray}
\label{e9}
L=\frac{1}{2}m_{ij}(x)\dot{x^i}\dot{x^j}-V(x)
\end{eqnarray} 
it was shown by Jacobi that the constrained motion of a particle with energy $E$ is provided by  geodesics of the rescaled  metric 
\begin{eqnarray}
\label{f3}
j_{ij}dx^idx^j=2(E-V)m_{ij}dx^idx^j   .
\end{eqnarray} 
It is interesting to observe that particle interactions can induce a curvature in the Jacobi metric through the potential function in an otherwise flat Newtonian space. One of the early workers in this topic was Pin \cite{ong} who considered many body systems  and in particular showed that the Gaussian curvature of the Jacobi metric has opposite sign to the particle energy $E$ (non-relativistic, without the rest energy). In later times Gibbons and coworkers \cite{gib3,gib4} have considered the optical metric in various physical situations which is a closely related concept for massless particles. Discussions on Jacobi metric approach in modern perspective can be found in \cite{deri}.

Again very recently the Jacobi metric formalism in relativistic scenario has been applied by Gibbons \cite{gib} in an elegant study of massive particle motion in Schwarzschild spacetime. In the present paper we closely follow and extend this work to massive particle motion in  Reissner - Nordstr\"{o}m  background. Indeed, Jacobi metric for a more general metric, that is the Kerr-Newman, has been derived in \cite{gibb} although the probe particle dynamics was not considered there. We consider both cases of the probe particle being neutral and charged. The results show a qualitative difference between the two cases since in the latter one needs to consider the additional Coulomb interaction term between the source and probe particle. For neutral particle the correction terms depend on $Q^2$, $Q$ being the charge of the black hole but interestingly for the charged probe correction terms involve $qQ$ terms as well, $q$ being charge of the particle, showing that the relative sign between the particle and black hole charge becomes important.

For the generic metric
\begin{eqnarray}
\label{a1}
ds^{2}=-V^{2} dt^2+g_{ij}dx^{i}dx^{j}
\end{eqnarray}	
 the action for a massive particle in this background can be written as,
\begin{eqnarray}
\label{a2}
S=-m\int Ldt=-m\int dt\sqrt{V^{2}-g_{ij}\dot{x^{i}}\dot{x^{j}}}.
\end{eqnarray}  
The canonical momentum 
\begin{eqnarray}
	\label{a3}
	p_i=\frac{m\dot{x^i}}{\sqrt{V^2-g_{ij}\dot{x^i}\dot{x^j}}},
\end{eqnarray}
leads to the Hamiltonian 

\begin{eqnarray}
	\label{a4}
	H=\sqrt{m^2V^2+V^2g^{ij}p_ip_j}.
\end{eqnarray}
This provides the Hamilton-Jacobi equation for the geodesics, parameterized by the energy $E$,
\begin{eqnarray}
\label{a5}
\sqrt{m^{2}V^{2}+V^{2}g^{ij}\partial^{i}S\partial_{j}S} = E
\end{eqnarray}
where  $p_i=\partial_i S$ .
Finally,  the  Hamiltonian-Jacobi equation for geodesics of Jacobi-metric $j_{ij}$ is given by,
 \begin{eqnarray}
\label{a6}
\frac{1}{E^{2}-m^{2}V^{2}}f^{ij}\partial_{i}S\partial_{j}S = 1
\end{eqnarray}
where $j_{ij}$ is defined as
\begin{eqnarray}
\label{a7}
j_{ij}dx^{i}dx^{j} = (E^{2} - m^{2}V^{2})V^{-2}g_{ij}dx^{i}dx^{j}.
\end{eqnarray}
Infact  $f_{ij}=V^{-2}g_{ij}$ turns out to be  the optical or Fermat metric. For massless particles $(m = 0)$ the Jacobi metric becomes equal to  the Fermat metric modulo a factor of $E^{2}$ and subsequently the geodesics do not depend upon energy E. However in the massive case, $m\neq 0$, the geodesics are $E$-dependent.

Let us put our  work in its proper perspective. The explicit results and observations of the present work are not entirely new.  Some of these are  discussed in the book by Chandrasekhar \cite{chan}. More recent and detailed studies are provided in   \cite{p1,p2,p3,oli}. However we have revisited the system from the Jacobi metric point of view. The Jacobi metric construction for charged massive particle and the subsequent analysis is completely new. Furthermore in the present work {\it{ the Gaussian curvature induced by the Jacobi metric plays an essential role since the motion is geodesic in nature}}. We have compared results computed from Gaussian curvature consideration (in Jacobi metric approach) with similar results obtained via conventional  scheme. For the charged probe,  the interplay between the gravitational and Coulomb forces proves to be interesting.

\section{Jacobi metric for  neutral particle in Reissner - Nordstr\"{o}m Geometry}
Reissner-Nordstr\"{o}m metric is a spherically symmetric solution of the coupled Maxwell Einstein gravity. It represents a black hole with a mass M and a charge Q and is given by 
\begin{eqnarray}
\label{a8}
ds^{2} =-(1-\frac{2M}{r}+\frac{Q^2}{r^2})c^2dt^2 +\frac{ dr^2}{(1-\frac{2M}{r}+\frac{Q^2}{r^2})}+r^2d\theta^2+r^2\sin^2\theta d\phi^2 .
\end{eqnarray}
For $Q=0$ the Schwarzschild metric is recovered.  As is well known it has two horizons at  $r=r_{\pm}=M \pm{\sqrt{M^2-Q^2}}$. The nature of the horizon singularities are different in Reissner - Nordstr\"{o}m  and Schwarzschild geometries.  The latter is spacelike whereas the former is timelike and thus yielding richer possibilities regarding the nature of trajectories. There are timelike worldlines for particles that can cross $r_+$ horizon and skirting the singularity can move out to another spacetime region after crossing $r_-$. On the contrary for Schwarzschild geometry, after crossing the event horizon at $r=2M$ the particle has no option but to fall towards the singularity. The $r_+$-horizon acts like the $r=2M$ event horizon of Schwarzschild whereas $r_-$-horizon is termed as the Cauchy horizon.

Now, generalizing the result of Gibbons \cite{gib},  the Jacobi metric corresponding to Reissner - Nordstr\"{o}m solution is
\begin{eqnarray}
\label{a9}
ds^2 = (E^2 - m^2 + \frac{2Mm^2}{r} - \frac{Q^{2}m^{2}}{r^{2}})[\frac{dr^{2}}{(1-\frac{2M}{r}+\frac{Q^{2}}{r^{2}})^{2}} + \frac{r^{2}d\theta^{2} + r^{2}\sin^{2}\theta d\phi^{2}}{(1-\frac{2M}{r}+\frac{Q^{2}}{r^{2}})}]
\end{eqnarray}
The first part is the conformal factor whereas the second factor is the optical metric. Due to  spherical symmetry, we are allowed to study the system in the  equatorial plane $ \theta = \frac{\pi}{2} $ without any loss of generality. This reduces the Jacobi metric to the form, 
\begin{eqnarray}
\label{a10}	
ds^{2} = (E^{2} - m^{2} + \frac{2Mm^{2}}{r} - \frac{Q^{2}m^{2}}{r^{2}})[\frac{dr^{2}}{(1-\frac{2M}{r}+\frac{Q^{2}}{r^{2}})^{2}} + \frac{r^{2} d\phi^{2}}{(1-\frac{2M}{r}+\frac{Q^{2}}{r^{2}})}]
\end{eqnarray}
Because of axial symmetry $\phi $ is a cyclic coordinate so that  the  angular momentum $l$ is conserved, 
  \begin{eqnarray}
 \label{b1}
 l = (E^{2} - m^{2} + \frac{2Mm^{2}}{r} - \frac{Q^{2}m^{2}}{r^{2}})(\frac{r^{2} }{(1-\frac{2M}{r}+\frac{Q^{2}}{r^{2}})})(\frac{d\phi}{ds})= constant .
\end{eqnarray}
Now, together with (\ref{a10}), (\ref{b1}) yields
\begin{eqnarray}
\label{b2}
(E^{2} - m^{2} + \frac{2Mm^{2}}{r} - \frac{Q^{2}m^{2}}{r^{2}})[\frac{1}{(1-\frac{2M}{r}+\frac{Q^{2}}{r^{2}})^{2}}(\frac{dr}{ds})^{2} +\frac{r^{2}}{(1-\frac{2M}{r}+\frac{Q^{2}}{r^{2}})}(\frac{d\phi}{ds})^{2} ] = 1,
\end{eqnarray}
that can be rewritten as
\begin{eqnarray}
\label{b3}
(E^{2} - m^{2} + \frac{2Mm^{2}}{r} - \frac{Q^{2}m^{2}}{r^{2}})^{2}\frac{1}{(1-\frac{2M}{r}+\frac{Q^{2}}{r^{2}})^{2}}(\frac{dr}{ds})^{2} = E^{2} - (1-\frac{2M}{r}+\frac{Q^{2}}{r^{2}})(m^{2} +\frac {l^{2}}{r^{2}}).
\end{eqnarray}
This  satisfies the standard result,
\begin{eqnarray}
\label{b4}
m^{2}(\frac{dr}{d\tau})^{2} =  E^{2} - (1-\frac{2M}{r}+\frac{Q^{2}}{r^{2}})(m^{2} +\frac {l^{2}}{r^{2}}).
\end{eqnarray}
Here  $\tau$ is the proper time along geodesic  and
 the angular momentum is
 \begin{eqnarray}
 \label{b5}
 l = mr^{2}(\frac{d\phi}{d\tau})
 \end{eqnarray}
 for
 \begin{eqnarray}
 \label{b6}
 d\tau = m\frac{(1-\frac{2M}{r}+\frac{Q^{2}}{r^{2}})}{(E^{2} - m^{2} + \frac{2Mm^{2}}{r}-\frac{Q^{2}m^{2}}{r^{2}})}ds.
\end{eqnarray}
Incidentally the above relation connects the proper time $\tau $ to the Jacobi path length $s$. Conventionally the trajectory is expressed in terms of a new variable $u=\frac{1}{r}$:
\begin{eqnarray}
\label{e7}
\frac{d^2u}{d\phi^2}+u=\frac{F(u)}{h^2u^2}
\end{eqnarray}
where, for the Reissner-Nordstr\"{o}m case, we find,
\begin{eqnarray}
\label{e8}
\frac{F(u)}{h^2u^2}=3Mu^2+\frac{M}{h^2}-2Q^2u^3-\frac{Q^2}{h^2}u.
\end{eqnarray}
In the above $h = l/m $ is the conserved angular momentum per unit mass. Thus (\ref{e7},\ref{e8}) constitute  the particle worldline or  trajectory equation.

It is straightforward to solve  (\ref{e7}) and generate the following first order differential equation,
\begin{eqnarray}
\label{b11}
(\frac{du}{d\phi})^{2} = -Q^{2}u^{4} + 2Mu^{3} - u^{2}(1 + \frac{Q^{2}}{h^{2}}) + \frac{2M}{h^{2}}u + C 
=-Q^2(u-\alpha)(u-\beta)(u-\gamma)(u-\delta)
\end{eqnarray}
 where, C is a constant related to the energy per unit mass $ \in = \frac{E}{m}$ by
\begin{eqnarray}
\label{b8}
C  = \frac{\in^{2} - 1}{h^{2}}.
\end{eqnarray}
This constitutes the first integral of motion and is one of our major results. Incidentally $\frac{h}{\in}$ is the impact parameter.  We immediately notice a qualitative change the black hole charge has brought about in the trajectory. In comparison to the Schwarzshild case \cite{gib} the present result involves a quartic term in $u$. Writing the quartic polynomial in terms of roots,
\begin{eqnarray}
\label{b7}
(\frac{du}{d\phi})^{2}=-Q^2(u-\alpha)(u-\beta)(u-\gamma)(u-\delta),
\end{eqnarray}
the following identities are recovered,
\begin{eqnarray}
\label{f7}
\alpha+\beta+\gamma+\delta=\frac{2M}{Q^2}, ~ ~~
\alpha\beta+\beta\gamma+\gamma\delta+\alpha\gamma+\alpha\delta+\beta\delta=\frac{1+\frac{Q^2}{h^2}}{Q^2},\\  ~~\alpha\beta\gamma\delta=-\frac{C}{Q^2},~~
~\alpha\beta\gamma+\beta\gamma\delta+\gamma\delta\alpha+\alpha\beta\delta=\frac{2M}{Q^2}{h^2}.
\end{eqnarray}
Clearly these are extensions of analogous relations given in \cite{gib} for the Schwarzshild metric.

Our aim is to propose explicit solutions for the trajectories in the same manner as those derived  in \cite{gib}. To that end let us quickly recapitulate  the orbit equation  for  Schwarzschild black hole \cite{hill,gib},
\begin{eqnarray}
\label{c1}
(\frac{du_G}{d\phi})^{2} =   2Mu_G^{3} - u_G^{2}+ \frac{2M}{h^{2}}u_G + C_G
\end{eqnarray}
with the explicit solution,
\begin{eqnarray}
\label{b9}
u_{G} = A_G + \frac{B_G}{\cosh^{2}(\omega_G\phi)}.
\end{eqnarray}
The constant parameters $A_G, B_g,C_G, \omega_G$  are given in appendix, with the subscript $G$ standing for the Gibbons solution \cite{gib}.

Let us now come to our work.
An explicit  solution for the trajectory equation for Reissner - Nordstr\"{o}m case (\ref{b11}) studied here is given by
\begin{eqnarray}
\label{c2}
u = A + \frac{B}{\cosh^{2}(\omega\phi)} + Q^{2}\frac{k}{\cosh^4(\omega\phi)}
\end{eqnarray}
where $ A, B, k $ all are constants. One can  solve for the constants perturbatively for small $Q$ to first non-trivial order in $Q^2$. The result is given in Appendix A. This constitutes one of our new results. This is a generic form of closed orbit. As we have pointed out in the Introduction, this type of analytic result for a generic orbit, indeed in an  approximate sense of of small charge $Q$ on the black hole) can compliment the graphical analysis of \cite{p1,p2} without any approximations.

It is straightforward to reduce our analysis to circular orbit  orbits for which $u=u_c$, a constant. We rewrite (\ref{b11}) as,
\begin{eqnarray}
\label{z1}
(\frac{du}{d\phi})^{2} = -Q^{2}u^{4} + 2Mu^{3} - u^{2}(1 + \frac{Q^{2}}{h^{2}}) + \frac{2M}{h^{2}}u + \frac{\epsilon^2-1}{h^2}
\equiv f(u) .
\end{eqnarray}
Presence of a biquadratic term of $u$ in $f(u)$, compared to a cubic one in  Schwarzschild geometry, is the qualitative change that leads to  significant  difference only for the orbits that cross the event horizon at $r_+$ and can skirt the singularity and come out of  $r=r_+$, (as discussed earlier), instead of  terminating at the singularity at $r=0$ as in the case of Schwarzschild black hole (see for example \cite{chan} for details).

For the occurrence of circular orbits  the conditions are as follows,
\begin{eqnarray}
\label{z2}
f(u) = -Q^{2}u^{4} + 2Mu^{3} - u^{2}(1 + \frac{Q^{2}}{h^{2}}) + \frac{2M}{h^{2}}u + \frac{\epsilon^2-1}{h^2}
=0,
\end{eqnarray}
\begin{eqnarray}
\label{z3}
f' (u) = -4Q^{2}u^{3} + 6Mu^{2} - 2u(1 + \frac{Q^{2}}{h^{2}}) + \frac{2M}{h^{2}} 
=0.
\end{eqnarray}
The relations change for the null geodesics which we are not considering at present (see for example \cite{chan}).
We can easily obtain the expressions for energy and angular momentum of a circular orbit of radius $r_c=\frac{1}{u_c}$ from the above two equations.
The expressions are,
\begin{eqnarray}
\label{z4}
\epsilon^2=\frac{(1-2Mu_c+Q^2u_c^2)^2}{1-3Mu_c+2Q^2u_c^2}
\end{eqnarray}
and,
\begin{eqnarray}
\label{z5}
h^2=\frac{M-Q^2u_c}{u_c(1-3Mu_c+2Q^2u_c^2)}.
\end{eqnarray}
Incidentally, these results agree with \cite{p1,p2}.

The minimum radius for a stable circular orbit will occur at the point of inflection of the function $f(u)$, i.e.,
\begin{eqnarray}
\label{z6}
f '' (u)=-12Q^2u^2+12Mu-2(1+\frac{Q^2}{h^2})=0.
\end{eqnarray}
Eliminating $h^2$ from the above equation using (\ref{z5})  we obtain,
\begin{eqnarray}
\label{z7}
4Q^4u_c^3-9 M Q^2u_c^2+6M^2u_c-M=0,
\end{eqnarray}
or, in terms of $r_c$,
\begin{eqnarray}
\label{z8}
r_c^3-6Mr_c^2+9Q^2r_c-\frac{4Q^4}{M}=0.
\end{eqnarray}

From (\ref{z8}) , for $Q^2=0$ , we recover the well known result for  Schwarzschild geometry,   $r_c=6M$. However, neglecting $Q^4$, a leading order correction to the radius is easily obtained, 
\begin{eqnarray}
\label{z88}
r_c\approx 3M\pm 3M(1-\frac{Q^2}{M^2})^{1/2}\approx 6M-\frac{3}{2}\frac{Q^2}{M}.
\end{eqnarray}

Using (\ref{z2}), (\ref{z3}) and (\ref{z6}),   (\ref{z1}) takes the form,
\begin{eqnarray}
\label{z9}
(\frac{du}{d\phi})^2=(u-u_c)^3(2M-3Q^2u_c-Q^2u) ,
\end{eqnarray}
and the solution is given by,
\begin{eqnarray}
\label{z10}
u=u_c+\frac{2(M-2Q^2u_c)}{(M-2Q^2u_c)^2(\phi-\phi_0)^2+Q^2} .
\end{eqnarray}

For Reissner-Nordstrom case,
to  $O(\frac{Q^2}{M})$ we obtain
\begin{eqnarray}
	\label{x9}
	r_c=6M-\frac{3Q^2}{2M}.
\end{eqnarray}
 This constitutes the first part of our work.
 \section{Jacobi metric for  charged particle in Reissner - Nordstr\"{o}m Geometry}
 The next level of generalization is to consider the trajectory of a probe with charge $q$ in the presence of a charged black hole. Indeed this is a non-trivial extension to the previous case since an additional Coulomb interaction term of the form $\sim (qQ)/r$ is involved.
  In Reissner-Nordstrom geometry where a test particle has a  charge per unit mass $q$, the only non vanishing component of the vector potential is $A_0$ and its motion is determined by the Lagrangian as of the form,
  \begin{eqnarray}
  \label{y1}
  2L=\left[(1-\frac{2M}{r}+\frac{Q^2}{r^2})(\frac{dt}{d\tau})^2-\frac{1}{(1-\frac{2M}{r}+\frac{Q^2}{r^2})}(\frac{dr}{d\tau})^2-r^2(\frac{d\theta}{d\tau})^2-r^2(\sin^2\theta)(\frac{d\phi}{d\tau})^2\right]+2\frac{qQ}{r}\frac{dt}{d\tau}
  \end{eqnarray}
 However, we need not attempt to construct  a generalized Reissner - Nordstr\"{o}m solution starting from the Einstein-Maxwell point charge action. The Jacobi metric formalism provides a quick answer.  Thus for the  Reissner - Nordstr\"{o}m case with a charged probe, the Jacobi metric is given by
 \begin{eqnarray}
 \label{y2}
 ds^2 = \left((E-\frac{mqQ}{r})^2 - m^2 + \frac{2Mm^2}{r} - \frac{Q^{2}m^{2}}{r^{2}}\right)\left[\frac{dr^{2}}{(1-\frac{2M}{r}+\frac{Q^{2}}{r^{2}})^{2}} + \frac{r^{2}d\theta^{2} + r^{2}\sin^{2}\theta d\phi^{2}}{(1-\frac{2M}{r}+\frac{Q^{2}}{r^{2}})}\right]
 \end{eqnarray}
 
 The only distinct feature which arises due to the probe charge $q$ in this case is that the energy for a particle (having a turning point) that arrives at the event horizon will be ,
 $E=\frac{mqQ}{r_+}$  
 and this can be   negative if $qQ<0$ which gives rise to the theoretical speculation of generating energy from a black hole. Once again a restriction of the motion to the equatorial plane,  i.e. $\theta=\frac{\pi}{2}$, reduces the Jacobi metric to  
\begin{eqnarray}
\label{e5}
ds^2 = \left((E-\frac{mqQ}{r})^2 - m^2 + \frac{2Mm^2}{r} - \frac{Q^{2}m^{2}}{r^{2}}\right)\left[\frac{dr^{2}}{(1-\frac{2M}{r}+\frac{Q^{2}}{r^{2}})^{2}} + \frac{ r^{2} d\phi^{2}}{(1-\frac{2M}{r}+\frac{Q^{2}}{r^{2}})}\right].
\end{eqnarray}

In an identical fashion, as done in previous cases, we derive the trajectory equation for $u=\frac{1}{r}$,
\begin{eqnarray}
\label{c6}
(\frac{du}{d\phi})^{2} = -Q^{2}u^{4} + 2Mu^{3} - u^{2}(1+\frac{Q^2}{h^2}-\frac{Q^2q^2}{h^2}) + \frac{2}{h^{2}}(M-\frac{qQE}{m})u + C
\end{eqnarray}
where $h = l/m $ is the conserved angular momentum per unit mass, $C$ is a constant related to the energy per unit mass $ \in = \frac{E}{m}$ and the angular momentum per unit mass $h$ is by the relation
$C  = \frac{\in^{2} - 1}{h^{2}}$ (as defined earlier in (\ref{b8})).

Again we discuss our work of constructing the orbit.
As we are considering  upto the first order correction terms of $Q^2$ so we can drop  the  $Q^2q^2$ term from (\ref{c6}) in our approximation.
The solution of (\ref{c6}) can be written as,
\begin{eqnarray}
\label{e1}
u = A_q + \frac{B_q}{\cosh^{2}(\omega_q\phi)} + Q^{2}\frac{k_q}{\cosh^4(\omega_q\phi )}
\end{eqnarray}
where $ A_q, B_q, k_q, \omega_q $ all are constant and their approximate expressions are once again provided in the Appendix B.

The trajectory equation for the charged probe is,
\begin{eqnarray}
\label{x1}
(\frac{du}{d\phi})^{2} = -Q^{2}u^{4} + 2Mu^{3} - u^{2}(1 + \frac{Q^{2}}{h^{2}}-\frac{Q^2q^2}{h^2}) + \frac{2}{h^{2}}(M-\frac{qQE}{m})u + \frac{\epsilon^2-1}{h^2}
\equiv f(u) .
\end{eqnarray}
This is the other principal result of our paper.

We now concentrate on circular trajectories. For the occurrence of circular orbits the conditions are as follows,
\begin{eqnarray}
\label{x2}
f(u) = 	 -Q^{2}u^{4} + 2Mu^{3} - u^{2}(1 + \frac{Q^{2}}{h^{2}}-\frac{Q^2q^2}{h^2}) + \frac{2}{h^{2}}(M-\frac{qQE}{m})u + \frac{\epsilon^2-1}{h^2}
=0,
\end{eqnarray}
and
\begin{eqnarray}
\label{x3}
f^\prime (u) = -4Q^{2}u^{3} + 6Mu^{2} - 2u(1 + \frac{Q^{2}}{h^{2}}-\frac{Q^2q^2}{h^2}) + \frac{2}{h^{2}}(M-\frac{qQE}{m}) 
=0.
\end{eqnarray}

We can easily obtain the expressions for energy and angular momentum of a circular orbit of radius $r_c=\frac{1}{u_c}$ from the above two equations.
The expressions for energy is,
\begin{eqnarray}
\label{x4}
\epsilon^2=\frac{E^2}{m^2}=\frac{(1-2Mu_c+Q^2u_c^2)^2+qQu_c[\frac{E}{m}(1-4Mu_c+3Q^2u_c^2)+ qQ u_c^2(M-Q^2u_c)]}{1-3Mu_c+2Q^2u_c^2}.
\end{eqnarray}
Now, upto  order $qQ$  (ignoring the term $q^2Q^2$ ), the expression can be written as,
\begin{eqnarray}
\label{t1}
\epsilon^2=\frac{(1-2Mu_c+Q^2u_c^2)^2}{(1-3Mu_c+2Q^2u_c^2)} + qQ u_c\left[\frac{(1-4Mu_c+3Q^2u_c^2)(1-2Mu_c+Q^2u_c^2)}{(1-3Mu_c+2Q^2u_c^2)^\frac{3}{2}} \right].
\end{eqnarray}
Apart from some minor  mismatch in numerical factors this result essentially agrees with \cite{p1,p2}.
Similarly for angular momentum, we have
\begin{eqnarray}
\label{x5}
h^2=\frac{(M-Q^2u_c)-qQ(\frac{E}{m}-qQu_c)}{u_c(1-3Mu_c+2Q^2u_c^2)}.
\end{eqnarray}
Thus, upto order of $qQ$ ,
\begin{eqnarray}
\label{t2}
h^2=\frac{(M-Q^2u_c)}{u_c(1-3Mu_c+2Q^2u_c^2)}-qQ \left[ \frac{(1-2Mu_c+Q^2u_c^2)}{u_c(1-3Mu_c+2Q^2u_c^2)^\frac{3}{2}}\right].
\end{eqnarray}
There is an interesting observation regarding a possible scaling of the charges following \cite{chan} where variations of $\epsilon ^2 =(E/m)^2$ and $h=l/m$ against $Mu_c$ are discussed with the scaling $Q^2=pM^2$, $p$ being a numerical constant. The resulting relation for the latter for neutral probe is \cite{chan}
\begin{eqnarray}
	\label{t3}
	\frac{h^2}{M^2}=\frac{(1-pMu_c)}{Mu_c(1-3Mu_c+2pM^2u_c^2)}.
\end{eqnarray}
However if we consider an identical scaling in our present case of with a charged probe, the relation turns out to be,
\begin{eqnarray}
	\label{t4}
	\frac{h^2}{M^2}=\frac{(1-pMu_c)}{Mu_c(1-3Mu_c+2pM^2u_c^2)}-q \sqrt{p} \left[ \frac{(1-2Mu_c+pM^2u_c^2)}{Mu_c(1-3Mu_c+2pM^2u_c^2)^\frac{3}{2}}\right].
\end{eqnarray}
Appearance of the parameter $q$ is indicative of the fact that the Coulomb force is essentially non-geometric and hence the trajectories are not pure geodesic in nature.

The effect of the probe charge, especially whether it is of same or opposite sign as the black hole charge, is quite striking. Intuitively we can argue that for the opposite sign case the probe charge effect will not be very significant because both the gravitational force and Coulomb force will be attractive and so qualitatively similar behavior to the neutral case will be observed. This is shown in Figure 1.

On the other hand, if the probe and black hole charges are of same sign the Coulomb force will be repulsive whereas the gravitational force is attractive as before. Interplay between   these two forces produces an upper bound of the $q$ parameter above which the results become unphysical. This is demonstrated in Figure 2.

Similar behavior for $\epsilon ^2=(E/m)^2$ vs. $r_c/M$  is observed in Figure 3 and in Figure 4 where, negative and positive values of $q$ are considered respectively.

\begin{figure}[htb!]
	{\centerline{\includegraphics[width=9cm, height=6cm] {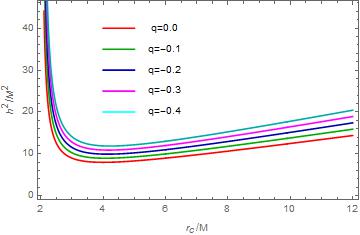}}}
	\caption{$h^2/M^2$ vs. $r_c/M$ are plotted for fixed  $Q^2=pM^2,~p=1$ and different negative values of $q$. The curves of charged probes are always above the neutral probe but of same qualitative nature.} \label{fig1}
\end{figure}
\vspace {.5cm}
\begin{figure}[htb!]
	{\centerline{\includegraphics[width=9cm, height=6cm] {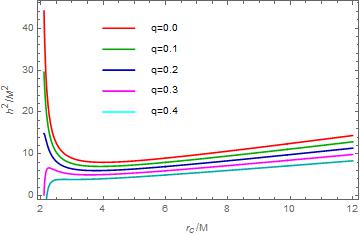}}}
	\caption{$h^2/M^2$ vs. $r_c/M$ are plotted for fixed  $Q^2=pM^2,~p=1$ and different positive values of $q$. The curves of charged probes are always below the neutral probe. The nature of curve changes for $q ~>0.2$ and becomes unphysical at $\sim 0.3$ onwards.} \label{fig2}
\end{figure}

\begin{figure}[htb!]
	{\centerline{\includegraphics[width=9cm, height=6cm] {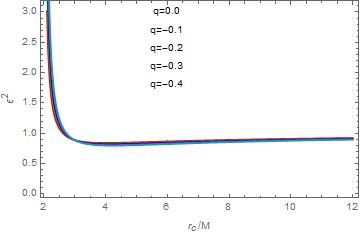}}}
	\caption{$\epsilon^2$ vs. $r_c/M$ are plotted for fixed  $Q^2=pM^2,~p=1$ and different negative values of $q$. } \label{fig3}
\end{figure}
\begin{figure}[htb!]
	{\centerline{\includegraphics[width=9cm, height=6cm] {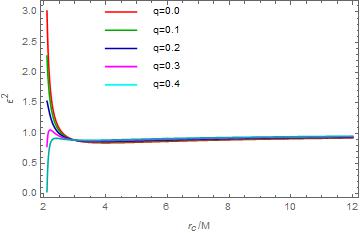}}}
	\caption{$\epsilon^2$ vs. $r_c/M$ are plotted for fixed  $Q^2=pM^2,~p=1$ and different positive values of $q$.} \label{fig4}
\end{figure}

The minimum radius for a stable circular orbit will occur at the point of inflection of the function $f(u)$, i.e.,
\begin{eqnarray}
\label{x6}
f''(u)=-12Q^2u^2+12Mu-2(1+\frac{Q^2}{h^2}-\frac{Q^2q^2}{h^2})
\end{eqnarray}
Eliminating $h^2$ from the above equation using (\ref{x5}) , we can write,
\begin{eqnarray}
\label{x7}
4Q^4u_c^3-9 M Q^2u_c^2+6M^2u_c-M-qQ \left[qQ(4u_c^3Q^2-3Mu_c^2)-\frac{E}{m}(6Q^2u_c^2-6Mu_c+1)\right]=0,
\end{eqnarray}
or, in terms of $r_c$ ,
\begin{eqnarray}
\label{x8}
r_c^3-6Mr_c^2+9Q^2r_c-\frac{4Q^4}{M}-\frac{qQE}{m}(\frac{r_c^3}{M}-6r_c^2+\frac{6r_cQ^2}{M})-q^2Q^2(3r_c-\frac{4Q^2}{M})=0.
\end{eqnarray}
Let us define a parameter $\Lambda=(qQE)/(mM) $, in terms of which the above equation is rewritten as
\begin{eqnarray}
\label{x8x}
r_c^3(1-\Lambda )-6Mr_c^2(1-\Lambda )+9Q^2r_c(1-\frac{2}{3}\Lambda -\frac{q^2}{3})-\frac{4Q^4}{M}(1-q^2)=0 .
\end{eqnarray}
Ignoring $O(Q^4)$ terms we obtain
\begin{eqnarray}
	\label{x10}
	r_c=3M\pm 3M{\sqrt{1-\frac{Q^2(1-\frac{2}{3}\Lambda -\frac{q^2}{3})}{M^2(1-\Lambda )}}}.
\end{eqnarray}
Note that for $q=0$ that is neutral probe the earlier result (\ref{x9}) $$r_c=6M-\frac{3Q^2}{2M}+...$$ is recovered. Considering $q$ and $\Lambda $ to be small (which is quite natural) we find
\begin{eqnarray}
\label{xxx10}
r_c=6M- \frac{3Q^2}{2M}(1+\frac{\Lambda}{3}).
\end{eqnarray}
This shows that the effective charge of the black hole increases for positive $\Lambda $ that is when the probe charge  $q$ and black hole charge $Q$ are of same sign but it decreases when they are of opposite sign. This concludes our discussion on charged particle trajectory in the presence of a charged black hole.

\section{Gaussian curvature for Jacobi metric}

Study of the Gaussian curvature for the Jacobi metric for Reissner- Nordstrom black hole, as experienced by a particle of fixed energy $E$ is the other major contribution of our work. 
The Jacobi metric corresponding to Reissner- Nordstrom geometry (\ref{a10}) (considering the system in the equatorial plane  $\theta=\frac{\pi}{2}$)
$$	
ds^{2} = (E^{2} - m^{2} + \frac{2Mm^{2}}{r} - \frac{Q^{2}m^{2}}{r^{2}})[\frac{dr^{2}}{(1-\frac{2M}{r}+\frac{Q^{2}}{r^{2}})^{2}} + \frac{r^{2} d\phi^{2}}{(1-\frac{2M}{r}+\frac{Q^{2}}{r^{2}})}]
$$
 vanishes when
\begin{eqnarray}
E^2=m^2(1-\frac{2M}{r}+\frac{Q^2}{r^2}).
\label{n2}
\end{eqnarray}
The equation saturates for a critical value $r_c$:
\begin{eqnarray}
r_c=\frac{Mm^2\left(1+(1-\frac{(m^2-E^2)Q^2}{M^2m^2})^{\frac{1}{2}}\right)}{m^2-E^2}\approx 2M(\frac{m^2}{m^2-E^2}-\frac{Q^2}{4M^2})+~O(Q^4).
\label{n3}
\end{eqnarray}
This leads to an inequality of the form $\frac{m^2}{m^2-E^2}\geq\frac{Q^2}{4M^2}$ and since $\frac{m^2}{m^2-E^2}<1$, one  has $Q<2M$. It is interesting to recall that $Q=M$ is the extremality condition and moreover $Q<M$ is generally assumed to avoid the presence of a naked singularity (or violation of Cosmic Sensorship hypothesis by Penrose). Hence  $Q<2M$ derived above is a weaker condition and does not add any further restrictions on the charge mass ratio.

 Interpreting the following expression as  Jacobi circumference \cite{gib} 
\begin{eqnarray}
2\pi(E^{2} - m^{2} + \frac{2Mm^{2}}{r} - \frac{Q^{2}m^{2}}{r^{2}})^{\frac{1}{2}}\frac{r}{(1-\frac{2M}{r}+\frac{Q^2}{r^2})^{\frac{1}{2}}}
\label{n4}
\end{eqnarray}
the boundary $r_c$ actually reduces to a point since the circumference vanishes there.

Condition for circular geodesics is derived from  the extrema of the Jacobi circumference,
\begin{eqnarray}
\frac{E^2}{m^2}=\frac{(1-\frac{2M}{r}+\frac{Q^2}{r^2})^2}{(1-\frac{3M}{r}+\frac{2Q^2}{r^2})}.
\label{n5}
\end{eqnarray}
Circular orbits exist provided $(1-\frac{3M}{r}+\frac{2Q^2}{r^2})\ge 0$. This means that to $O(Q^2/M)$ the roots are $$r_+=3M-(2Q^2)/(3M),~r_-=(2Q^2)/(3M)$$
indicating that $r\ge 3M-(2Q^2)/(3M)$.

Several examples of radii (to $O(Q^2)$) for some specific values of energy are provided below:
for $m^2=0$, the term $(1-\frac{3M}{r}+\frac{2Q^2}{r^2})$ must be equal to zero leading to  $r\approx 3M-\frac{2Q^2}{3M}$. The outermost and innermost circular orbit radii are computed by extremizing r.h.s of (\ref{n5}). These are respectively  $r\approx 6M-\frac{3Q^2}{2M}$ for $E^2\approx  m^2\left[\frac{8}{9}-\frac{2Q^2}{81M^2}\right]$ and $r\approx 4M-\frac{Q^2}{M}$ for $E^2=m^2$. Incidentally the last example agrees with our previously derived result  (\ref{x9}). For $Q=0$ the above  reduce to the Schwarschild geometry results \cite{gib}.

Let us now come to the explicit structure of Gaussian curvature $K$. As a warmup exercise let us compute $K_S$ for  Schwarschild case. Expressing the  Jacobi metric pertaining to  Schwarschild geometry as,
\begin{eqnarray}
\label{n6}
ds^2=f(r)^2[dr^2+r^2(1-\frac{2M}{r})d\phi^2]
\end{eqnarray}
where, $$f(r)^2=(E^2-m^2+\frac{2Mm^2}{r})\frac{1}{(1-\frac{2M}{r})^2}$$
the Gaussian curvature  $K_S$ is given by, 
\begin{eqnarray}
\label{n7}
K_S &=& -\frac{1}{f(r)^2r(1-\frac{2M}{r})^{\frac{1}{2}}}\frac{d}{dr}\left[\frac{1}{f(r)}\frac{d}{dr}\left(rf(r)(1-\frac{2M}{r})^{\frac{1}{2}}\right)\right] \\
&=& \frac{M[m^4(2M-r)^3+E^4r^2(3M-2r)+3E^2m^2r(2M-r)(M-r)]}{r^3[m^2(2M-r)+E^2r]^3}.
\end{eqnarray}
For massless probe $m=0$,  the curvature simplifies to,
\begin{eqnarray}
\label{n8}
K_S=-\frac{2M}{E^2r^3}(1-\frac{3M}{2r}).
\end{eqnarray}
For $r\geq 2M$, $K_S$ is always negative \cite{gib3}.
\vskip .3cm

{\it{Uncharged probe}}:\\
Let us start with massless neutral probe. 
In a similar way as described above, for the Reissner-Nordstrom case, the Gaussian curvature to $O(Q^2)$ is given by,
\begin{eqnarray}
\label{n91}
K_{RN}=K_S+(A/B)Q^2
\end{eqnarray}
where
\begin{eqnarray}
\label{n9}
A&=& 3E^6r^3(-2M+r)-m^6(-2M+r)^4+E^4m^2r^2(-16M^2+22Mr-7r^2) \nonumber \\
&& +E^2m^4r(-28M^3+42M^2r-24Mr^2+5r^3), \nonumber \\
 B&=&r^4[m^2(2M-r)+E^2r]^4
\end{eqnarray} 
For $m=0$ massless probe,  the Gaussian curvature reduces to the form,
\begin{eqnarray}
\label{n10}
K_{RN}=-\frac{1}{E^2}\left[\frac{2M}{r^3}(1-\frac{3M}{2r})-\frac{3Q^2}{r^4}(1-\frac{2M}{r})\right].
\end{eqnarray}

Comparing with $Q=0$ case (\ref{n8}) , the situation becomes more complicated and indeed, it is possible that the $Q$-contribution might reverse the sign of $K_{RN}$. To verify this we define a convenient scaling $E^2=c_1 M^2 , r=aM, Q=bM$ and   rewrite $K_{RN}$ in (\ref{n10}) as,
\begin{eqnarray}
\label{s1}
K_{RN}=-\frac{1}{a^3c_1M^4}[2(1-\frac{3}{2a})-\frac{3b^2}{a}(1-\frac{2}{a})].
\end{eqnarray}	
To see the effect of $Q$ we plot $K_{RN}$ vs. $c_1=E^2/M^2$ for fixed values of   $b$ in Fig. 5,  near the outer boundary where $a=6-\frac{3}{2}b^2$.

\begin{figure}[htb!]
	{\centerline{\includegraphics[width=9cm, height=6cm] {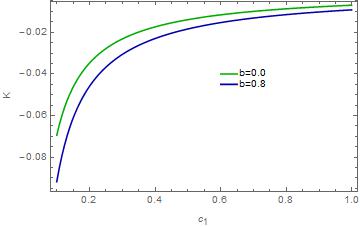}}}
	\caption{$K_{RN}$ vs. $c_1=E^2/M^2$ are plotted for $Q=bM$ where $b$ takes different positive values near the outer boundary.}  
	\label{fig5}
\end{figure}
The graphs indicate that, to the order of approximation we are considering, for massless probe, $K_{RN}$ stays negative, as in Schwarschild case \cite{gib} and as the energy of the particle increases, the magnitude of the Gaussian curvature decreases and asymptotically tends to become zero. Thus for massless probe, the presence of black hole charge has not too much effect except it increases the negativity of the Gaussian curvature.

Let us now consider a probe with mass $m$ where the dynamics changes qualitatively from the massless probe discussed above. We introduce the parameterization $E^2=c m^2 , r=aM, Q=bM$ and plot  $K_{RN}$ vs. $c=E^2/m^2$  (Fig.s 6,7,8). The Gaussian curvature, (to order  $Q^2$),  near outer boundary  $a=6-\frac{3}{2}b^2$, is
\begin{eqnarray}
\label{n11}
K_{RN}=\frac{-64 + 360 c - 324 c^2}{216 (-4 + 6 c)^3 m^2M^2}+\frac{(28 - 201 c + 333 c^2 - 162 c^3) b^2}{648 (-2 + 3 c)^4 m^2M^2}.
\end{eqnarray}
From Fig. 6 we notice that, for the energy range 
$0.7<E^2/m^2<0.8 $, for small $b^2$ (small $Q$), the Gaussian curvature $K_{RN}$ is positive but it changes sign and becomes negative for larger $Q\sim 0.5$  and from Fig. 7 we see that the transition occurs around $b^2=0.3$. Interestingly in this case the $K_{RN}$ starts with a negative value and briefly reaches positive values before vanishing asymptotically. From Fig.8, we find that for $c=E^2/m^2 \le 0.2$, $K_{RN}$ remains positive both for $Q=0$ and for non-zero $Q$.

For inner boundary, we put $a=4-b^2$ and $K_{RN}$ is 
\begin{eqnarray}
\label{n12}
K_{RN}=\frac{-8 + 72 c - 80 c^2}{64 (-2 + 4 c)^3 m^2M^2}+\frac{(2 - 23 c + 44 c^2 - 24 c^3) b^2}{256 (-1 + 2 c)^4 m^2M^2}.
\end{eqnarray}
We have chosen the same value $b^2=0.3$ for which the sign change occurred at outer boundary. The plot in Fig. 9 indicates that, near inner boundary,  for $b^2=0.3$, $K_{RN}$ tends to become more negative but the graphs shows similar qualitative nature for both zero and non zero $Q$.
\vskip .3cm

\begin{figure}[htb!]
	{\centerline{\includegraphics[width=9cm, height=6cm] {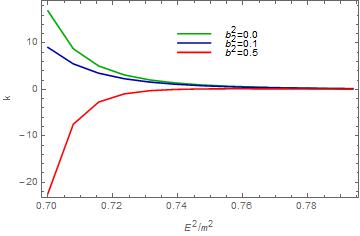}}}
	\caption{$K_{RN}$ vs. $c=E^2/m^2$ are plotted for $Q^2=b^2M^2$ and different positive values of $ b^2$ near the outer boundary $(0.7<E^2/m^2<0.8) $ } \label{fig6}
\end{figure}

\begin{figure}[htb!]
	{\centerline{\includegraphics[width=9cm, height=6cm] {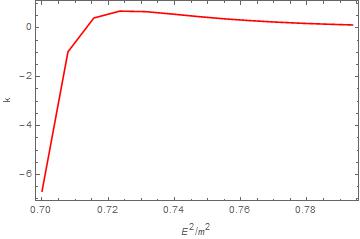}}}
	\caption{$K_{RN}$ vs. $c=E^2/m^2$ are plotted for $Q^2=b^2M^2$ where $ b^2=0.3$ near the outer boundary $(0.7<E^2/m^2<0.8) $ } \label{fig7}
\end{figure}

\begin{figure}[htb!]
	{\centerline{\includegraphics[width=9cm, height=6cm] {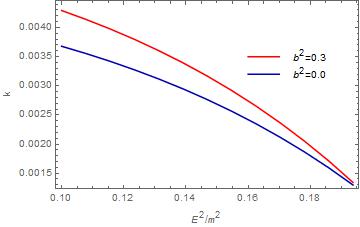}}}
	\caption{$K_{RN}$ vs. $c=E^2/m^2$ are plotted for $Q^2=b^2M^2$ and different positive values of $ b^2$ near the outer boundary $(E^2/m^2<0.2) $ } \label{fig8}
\end{figure}

\begin{figure}[htb!]
	{\centerline{\includegraphics[width=9cm, height=6cm] {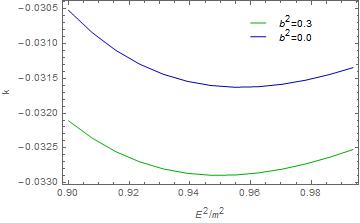}}}
	\caption{$K_{RN}$ vs. $c=E^2/m^2$ are plotted for $Q^2=b^2M^2$ and different positive values of $ b^2$ near the inner boundary $(0.9<E^2/m^2<1) $ } \label{fig9}
\end{figure}
{\it{Charged probe}}:\\
We now discuss the case of charged probe where an additional Coulomb interaction needs to be taken into account. The Jacobi metric corresponding to Reissner Nordstrom case with a charged probe ($q$) (in the equatorial plane)	is 
\begin{eqnarray}
\label{p1}
ds^2 = \left((E-\frac{mqQ}{r})^2 - m^2 + \frac{2Mm^2}{r} - \frac{Q^{2}m^{2}}{r^{2}}\right)\left[\frac{dr^{2}}{(1-\frac{2M}{r}+\frac{Q^{2}}{r^{2}})^{2}} + \frac{ r^{2} d\phi^{2}}{(1-\frac{2M}{r}+\frac{Q^{2}}{r^{2}})}\right].
\end{eqnarray}	

The metric vanishes when
\begin{eqnarray}
E^2-\frac{2EmQq}{r}=m^2(1-\frac{2M}{r}+\frac{Q^2}{r^2}),
\label{p2}
\end{eqnarray}
where the $q^2Q^2$-term has been dropped for small $q$. This leads to,
\begin{eqnarray}
r_c=\frac{(Mm^2-EmQq)\left(1+(1-\frac{(m^2-E^2)Q^2}{(Mm-EQq)^2})^{\frac{1}{2}}\right)}{m^2-E^2}.
\label{p3}
\end{eqnarray}
Again this can be thought of as a point where the Jacobi circumference vanishes.
The circumference can be written from the metric as,
\begin{eqnarray}
2\pi(E^{2} - m^{2} -\frac{2EmQq}{r} + \frac{2Mm^{2}}{r} - \frac{Q^{2}m^{2}}{r^{2}})^{\frac{1}{2}}\frac{r}{(1-\frac{2M}{r}+\frac{Q^2}{r^2})^{\frac{1}{2}}}.
\label{p4}
\end{eqnarray}

Circular geodesics correspond to the extrema of the Jacobi circumference for which,
\begin{eqnarray}
\frac{E^2}{m^2}=\frac{(1-\frac{2M}{r}+\frac{Q^2}{r^2})^2}{(1-\frac{3M}{r}+\frac{2Q^2}{r^2})}  \pm \frac{Qq}{r}\left[\frac{(1-\frac{4M}{r}+\frac{3Q^2}{r^2})(1-\frac{2M}{r}+\frac{Q^2}{r^2})}{(1-\frac{3M}{r}+\frac{2Q^2}{r^2})^\frac{3}{2}}\right] 
\label{p5}
\end{eqnarray}	 

For the  Reissner Nordstrom geometry where the probe charge $q$ is present, the Gaussian curvature can be written as,
\begin{equation}
\label{grnp}
K_{RNP}=K_{S}+(C/D)qQ
\end{equation}
where
\begin{eqnarray}
\label{n13}
C&=&  Em[E^4r^2(12M^2-9Mr+r^2)+m^4(-2M+r)^2(4M^2-Mr+r^2)]  \nonumber \\
&& -2E^3m^3r(-4M^3+12M^2r-7Mr^2+r^3)\nonumber \\
D&=& r^3[m^2(2M-r)+E^2r]^4
\end{eqnarray} 
It is worthwhile to point out that the leading  correction term depends on $qQ$ that both linearly on $q$ and $Q$ which leads to interesting consequences. Firstly, the relative sign of black hole charge $Q$ and probe charge $q$ becomes important as it dictates the nature of Coulomb interaction, that is whether it is repulsive or attractive. Secondly, unlike the Reissner Nordstrom geometry with neutral probe, here we can drop $Q^2$-terms due to the presence of $qQ$-terms.

Now where the probe charge $q$ is present, we parametrize $Q=M/2$, $r=\frac{15}{8}pM$ and $E^2/m^2=1$ and find the Gaussian Curvature as,
\begin{eqnarray}
	\label{s2}
	K=-\frac{8(-32+45p)(2+q)}{3375M^2p^3}
\end{eqnarray} 
From the above expression we can easily observe that for positive values of probe charge $q$ or in the absence of the probe charge, the graphs will show similar qualitative nature (shown in fig.10) where as for negative values of probe charge i.e. for $(2+q)<0$ or, $q<-2$ , it will reverse the sign of Gaussian curvature $K$ and the curvature starts with negative value and tends to become zero. 

Similarly, if we plot $K$ vs. $c=E^2/m^2$ for negative and positive values of probe charge $q$ , the graphs shows similar qualitative nature for positive values of probe charge whereas it shows reverse nature for negative values of probe charge(shown in fig.11).

\begin{figure}[htb!]
	{\centerline{\includegraphics[width=9cm, height=6cm] {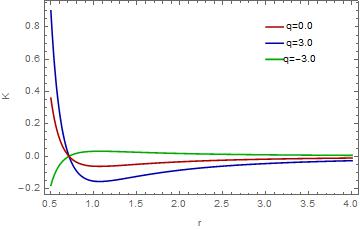}}}
	\caption{$K$ vs. $r$ are plotted for positive and negative values of probe charge $q$ for fixed $Q=M/2$.  } \label{fig10}
\end{figure}

 \begin{figure}[htb!]
 	{\centerline{\includegraphics[width=9cm, height=6cm] {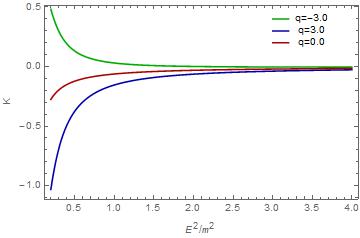}}}
 	\caption{$K$ vs. $c=E^2/m^2$ are plotted for positive and negative values of probe charge $q$ for fixed $Q=M/2$.  } \label{fig11}
 \end{figure}

\section{Conclusion} In the present work we have  considered particle trajectories that are parameterized by constant energy value. This feature helps to visualize quickly the bounded and unbounded nature of particle orbits related to particle energy. This characteristics is very succinctly incorporated in the Jacobi extension of least action principle. The formalism starts with the construction of the Jacobi metric where the (conserved) particle energy appears explicitly in the metric. As has been proved by Pin \cite{ong} a restricted variational principle $a~la$ Maupertuis with conventional metric and constant particle energy is equivalent to an unrestricted variational principle with Jacobi metric, which  explicitly involves the particle energy.  Hence the constant energy paths are still geodesics but of the Jacobi metric. 

Exploiting this formalism we have studied the worldlines of both   uncharged and charged probes in Reissner Nordstrom  background. The former is a straightforward generalization of the Schwarzschild black hole as given by Gibbons \cite{gib} whereas the latter is  a non-trivial extension since it has an additional Coulomb interaction. In both cases we have derived the circular orbit condition. The relative sign of the the probe and black hole (whether both have same sign or opposite sign) plays a significant role in determining nature of particle trajectory. Indeed, it should be stressed that there already exist series of recent \cite{p1} and earlier works \cite{p2,p3} by Pugliese, Quevedo and Ruffini where this problem has been treated in detail in an exact way from graphical perspective. We, on the other hand, have attempted to compliment their works by providing analytic form of the orbits in an approximate scenario of small charge for the black hole and probe. We have also given expressions for the generic closed orbit and its reduction to circular orbits.  

The Gauss curvature pertaining to Jacobi metric comprises another new and interesting aspect of our work. In earlier works  \cite{ong,gib}, Gaussian curvature of the Jacobi metric played an important role in characterizing the nature of the particle worldlines in terms of open or closed orbits related to the  particle energy. We have studied the properties of Gaussian curvature related to Jacobi metric for uncharged probe (massless and massive case) and also for the charged probe. 

 For massless uncharged probe, the presence of black hole charge $Q$ only increases the negativity of Gaussian curvaturen whereas for the massive case it plays a significant role i.e. the black hole charge can reverse the sign of Gaussian curvature for holes with charge-mass ratio above a certain value.
 
 When the probe is charged, an additional Coulomb interaction comes into play and the sign of  the probe charge,  whether it is of same sign or opposite to the black hole charge, becomes crucial. It is expected that  if the charges are of the same sign,   the effect becomes pronounced due to the competition between the repulsive  Coulomb force   and attractive gravitational force. On the other hand,  the effect will be weaker if the charges are of opposite sign since  both the Coulomb and gravitational forces will be attractive in nature. This is manifested in the angular momentum vs. circular radius and energy vs. circular radius graphs. However, curiously enough, the Gaussian curvature behaves in a different way.   For a positively charged black hole, if the probe charge is also positive,  the Gaussian curvature starts with a negative value and asymptotically tends to zero. This is similar to the case of a neutral probe. But if the probe has a negative charge ($q<-2$), (opposite to the positive black hole charge),   the curvature stays positive throughout and asymptotes to zero.

\vskip 1cm
{\bf{Acknowledgement:}}
It is a pleasure to thank Professor Gary Gibbons for many helpful  correspondences and for sending us reprints of early references. We are grateful to the referee for helpful comments that have helped to improve presentation of the paper. P.D. acknowledges the financial support from INSPIRE, DST, India.

\section{Appendix A}
Explicit solutions for the constants given in \cite{gib} for Schwarzschild background are,
\begin{equation}
\label{b10}
A_G=\frac{1}{6M}(1\pm \sqrt{1-\frac{12M^{2}}{h^{2}}})=\frac{1}{6M}(1-2MB_G)
\end{equation}
\begin{equation}
\label{x}
B_G=\mp\frac{1}{2M}\sqrt{1 - \frac{12M^{2}}{h^{2}}}
\end{equation}
\begin{equation}
\label{z}
\omega^{2}_G = \pm\frac{1}{4}\sqrt{1-\frac{12M^{2}}{h^{2}}}=-\frac{MB_G}{2}
\end{equation}
\begin{equation}
\label{y}
C_G = A^{2}_G(4MA_G - 1) =\frac{1}{36M^2}(1-2MB_G)^2\left[-\frac{4}{3}MB_G-\frac{1}{3}\right].
\end{equation}

Exploiting these  we provide below the  $O(Q^2)$ corrected expressions for the neutral probe  Reissner - Nordstr\"{o}m system, 
\begin{eqnarray}
A=A_G+Q^2 f, ~ B=B_G+Q^2 g, ~  \omega^2=\omega_G^2+Q^2 s,~
C=C_G+Q^2 t
\label{f1}
\end{eqnarray}
where, 
\begin{eqnarray}
\label{f2}
f=\frac{2A_G^3+\frac{A_G}{h^2}}{6MA_G-1}=-\frac{(1-2MB_G)}{2MB_G}\left[\frac{1}{108M^3}(1-2MB_G)^2+\frac{1}{6Mh^2}\right] ,\\
g=\frac{-2B_G s+4k\omega_G^2+2A_GB_G^2}{MB_G}=\frac{1}{MB_G}\left[-2B_G s\pm B_G^3+\frac{1}{3M}(1-2MB_G)B_G^2\right], \\
s=\frac{1}{4}\left[-6A_G^2\mp 3A_G^2+6Mf\pm \frac{A_G}{M}-\frac{1}{h^2}\mp \frac{1}{h^2} \right] \end{eqnarray}
or equivalently
\begin{eqnarray}
\label{ff2}
s=\frac{1}{4}\left[(-\frac{1}{6M^2}\mp \frac{1}{12M^2})(1-2MB_G)^2+6Mf\pm \frac{1}{6M^2}(1-2MB_G)-\frac{1}{h^2}\mp \frac{1}{h^2}\right], \\
t=A_G^2(A_G^2+\frac{1}{h^2})=\frac{1}{36M^2}(1-2MB_G)^2\left[\frac{1}{36M^2}(1-2MB_G)^2+\frac{1}{h^2}\right],  \\
k=\mp\frac{1}{8M^3}(1-\frac{12M^2}{h^2})=\mp \frac{B_G^2}{2M}.
\end{eqnarray}
\section{Appendix B}
Similarly,  the  $O(Q^2)$ corrected expressions for the charged probe  Reissner - Nordstr\"{o}m system are given by, 
\begin{eqnarray}
A_q=A_G+Q^2 f, ~ B_q=B_G+Q^2 g, ~  \omega_q^2=\omega_G^2+Q^2 s,~
C_q=C_G+Q^2 t
\label{e2}
\end{eqnarray}
but the changes appeared only for the expressions of $A_q$ and $C_q$.
\begin{eqnarray}
\label{e6}
A_q=A_G+Q^2\frac{(2MB_G-1)}{2MB_G}\left[\frac{1}{108M^3}(1-2MB_G)^2+\frac{1}{6Mh^2}\right]+Qq[\frac{E}{h^2(-2MB_G)}], \\
B_q=B_G+Q^2 \frac{1}{MB_G}\left[-2B_G s\pm B_G^3+\frac{1}{3M}(1-2MB_G)B_G^2\right],\\
\omega_q^2=\omega_G^2+\frac{Q^2}{4}\left[(-\frac{1}{6M^2}\mp \frac{1}{12M^2})(1-2MB_G)^2+6Mf\pm \frac{1}{6M^2}(1-2MB_G)-\frac{1}{h^2}\mp \frac{1}{h^2}\right], \\
C_q=C_G+Q^2 \frac{1}{36M^2}(1-2MB_G)^2\left[\frac{1}{36M^2}(1-2MB_G)^2+\frac{1}{h^2}\right]+Qq\left[\frac{E}{3Mh^2}(1-2MB_G)\right], \\
k_q=\mp\frac{B_G^2}{2M}
\end{eqnarray}

\end{document}